\documentclass[manuscript]{aastex6}

\usepackage{breakurl}
\usepackage{amsmath}
\usepackage{cases}

\shorttitle{ICME HEATING BY ALFV\'ENIC  FLUCTUATIONS DISSIPATION}
\shortauthors{LI ET AL.}

\begin{document}


\title{Plasma heating inside ICMEs by Alfv\'enic fluctuations dissipation}

\author{Hui Li \altaffilmark{1}, Chi Wang \altaffilmark{1}, Jiansen He \altaffilmark{2}, Lingqian Zhang \altaffilmark{1}, \\John D. Richardson \altaffilmark{3},  John W. Belcher \altaffilmark{3}, and Cui Tu \altaffilmark{4}}


\altaffiltext{1}{State Key Laboratory of Space Weather, National Space Science Center, CAS, Beijing, 100190, China; \url{hli@spaceweather.ac.cn}}

\altaffiltext{2}{School of Earth and Space Sciences, Peking University, Beijing, 100871, China}

\altaffiltext{3}{Kavli Institute for Astrophysics and Space Research, Massachusetts Institute of Technology, Cambridge, MA, USA}

\altaffiltext{4}{Laboratory of Near Space Environment, National Space Science Center, CAS, Beijing, 100190, China}



\begin{abstract}

Nonlinear cascade of low-frequency Alfv\'enic fluctuations (AFs) is regarded as one candidate of the energy sources to heat plasma during the non-adiabatic expansion of interplanetary coronal mass ejections (ICMEs). However, AFs inside ICMEs were seldom reported  in the literature. In this study, we investigate AFs inside ICMEs using observations from \textit{Voyager 2} between 1 and 6 au. It is found that AFs with high degree of Alfv\'enicity frequently occurred inside ICMEs, for almost all the identified ICMEs (30 out of 33 ICMEs), and 12.6\% of ICME time interval. As ICMEs expand and move outward, the percentage of AF duration decays linearly in general. The occurrence rate of AFs inside ICMEs is much less than that in ambient solar wind, especially within 4 au. AFs inside ICMEs are more frequently presented in the center and at the boundaries of ICMEs. In addition, the proton temperature inside ICME has a similar distribution. These findings suggest significant contribution of AFs on local plasma heating inside ICMEs.

\end{abstract}


\section{Introduction}
\label{sec:intro}
Coronal mass ejections (CMEs) are spectacular large-scale disturbed structures involving great explosion of solar material into heliosphere. Solar wind structures or interplanetary manifestations of CMEs are now generally referred to as interplanetary coronal mass ejections (ICMEs), which are the heliospheric counterparts of CMEs at the Sun \citep[e.g.][]{Gosling 1990,Neugebauer and Goldstein 1997}.

ICMEs often expand in size with radial distance in the inner heliosphere since their internal pressures are generally higher than the ambient solar wind, and their leading edges usually move faster than the trailing edges \citep[see][and references therein]{Burlaga 1995}. The radial width increases with distance out to $\sim$ 15 au (astronomical unit) \citep{Wang and Richardson 2004,Liu et al 2005}; beyond this distance the widths are relatively constant because ICMEs reach equilibrium with the background solar wind \citep{Wang et al 2005,Liu et al 2006,Richardson et al 2006}.

For an expanding ICME, the proton temperature would be expected to decrease more quickly within the ICME than in the ambient solar wind due to adiabatic cooling. However, the proton temperature inside ICMEs does not behave even qualitatively as expected. \citet{Liu et al 2005}  and \citet{Wang et al 2005} found that the proton temperature inside the ICMEs from 0.3 to 5.4 au decreases slower than in the background solar wind. \citet{Wang and Richardson 2004}, \citet{Richardson et al 2006} and \citet{Liu et al 2006} combined Voyager 1 and Voyager 2 data and extended such findings out to 30 au. The polytropic index $\gamma$ was determined empirically to be 1.15 $\sim$ 1.33, implying considerable local plasma heating within ICMEs.

The most probable energy source of plasma heating within ICMEs is believed to come from the magnetic field. However, the mechanism which dissipates magnetic energy into thermal energy is still an open question. In the literature, several candidate mechanisms were proposed to heat the CME plasma, which are listed as follows: (1) Outflows from the CME current sheets have the potential to heat CMEs far from the flare sites \citep{Bemporad et al 2007}; (2) Kinking is frequently observed in prominences and during solar eruptions \citep[e.g.,][]{Rust and LaBonte 2005}. The kink instability could heat CME plasma by injecting energy at outer scale for turbulence development through large-scale motions; (3) Heating by small-scale magnetic reconnection is another candidate mechanism and is similar to the nanoflare model of coronal heating. The frequently occurred tearing mode is one possible manifestation of this mechanism \citep{Furth et al 1963}; (4) The damping of MHD waves driven by photospheric motions is also one of the mechanisms of heating ICMEs; (5) Thermal conduction along magnetic field is quick and therefore regarded as a potential contributor to the heating of ICME plasma \citep{Landi et al 2010}; (6) Energetic particles accelerated during the impulsive phase of solar flares could contribute to CME heating; (7) Counteracting flows, such as upward concave flux rope segments yield shocks from colliding flows accelerated by gravity, may heat CME plasma \citep{Filippov and Koutchmy 2002}; and (8) Ohmic heating from net current in the flux rope (see \citet{Murphy et al 2011} and the references therein for more details).

Nonlinear cascade of low-frequency Alfv\'enic fluctuations (AFs), which transfers energy from large scales down to small kinetic scales for further dissipation, is generally regarded as one candidate  energy source for the heating of ICME plasma. It can preferentially heat heavy ions within ICMEs as observed \citep{Tu and Marsch 1995,Tam and Chang 1999,Kasper et al 2008,Wang et al 2014}. \citet{Galinsky and Shevchenko 2012} showed that heavy ions could be heated due to interactions between anti-sunward and sunward AFs. \citet{Liu et al 2006} found turbulence inside an ICME at 3.25 au and suggested that magnetic turbulence dissipation seems sufficient to explain the ICME heating; they assumed that the turbulence was driven by AFs, though AFs were seldom reported inside 280 ICMEs from 0.3 to 20 au. To our knowledge, quite limit numbers of  AF events haven been published in the literature. \citet{Marsch et al 2009} found possible AFs lasting for almost one hour in an ICME detected at 0.7 au. \citet{Yao et al 2010} later presented clear AFs with 2-hr duration inside an ICME observed at 0.3 au.

Traditional diagnosis of AFs may underestimate the degree of Alfv\'enicity, and thus possibly miss some AFs. \citet{Li et al 2016a} proposed a new approach to search for AFs, which could reduce the uncertainties in identifying AFs. In this study, we apply this AF diagnosis approach to identify AFs inside ICMEs from 1 to 6 au based on \textit{Voyager 2} data. Abundant AFs are found within ICMEs. Clear indirect evidence of the contributions of AFs on ICME plasma heating is provided.

\section{Diagnosis of AFs within ICMEs}
\label{sec:method}
The differences of ICMEs lie in the different signatures of magnetic field, plasma, composition and energetic particle. However, the identification of ICMEs still remains subjective undertaking. No single characteristic has proved both necessary and sufficient to define the presence of ICMEs. The currently used signatures for the in situ identification of ICMEs at 1 au have been well summarized \citep[see][and references therein]{Zurbuchen and Richardson 2006}, such as smooth magnetic field rotation and enhancement, low proton temperature, extreme density decrease, enhanced alpha/proton ratio, abundance and charge state anomalies of heavy ion species, bidirectional electron beams and cosmic rays, and so on. In the outer heliosphere, ICME identification becomes even more difficult. On one hand, some signatures of ICMEs are blurred through interaction with the ambient solar wind. On the other hand, current available measurements from the limited instruments cannot supply the complete set of variables required for comprehensive identification.

\citet{Wang and Richardson 2004} identified 147 probable ICMEs from hourly averaged solar wind plasma and magnetic field data by \textit{Voyager 2} between 1 and 30 au. The primary criterion they used is abnormally low solar wind proton temperature proposed by \citet{Richardson and Cane 1995}. This criterion compares the observed proton temperature $T_p$ with the ``expected" temperature $T_{ex}$ appropriate for ``normally expanding" with the observed solar wind speed $V$. $T_{ex}$ (in unit of $10^3$ K) is calculated from the relationship derived by \citet{Lopez 1987}:
\begin{linenomath}
\begin{equation}
\label{tex}
T_{ex} = \begin{cases}
(0.031V - 5.1)^2 / R^{0.6} & < 500 \text{ km/s} \\
(0.51 V - 142) / R^{0.6} & \geq 500 \text{ km/s}
\end{cases}
\end{equation}
\end{linenomath}
They also examined the magnetic field and plasma parameters to exclude regions which may not be ICMEs, such as regions associated with heliospheric current sheet crossings. The determination of ICME boundaries is uncertain since different signatures usually have different boundaries. The boundaries chose by them were generally coincident with the regions where $T_p / T_{exp} =  0.5$ with some adjustments based on reduced magnetic field fluctuations.

In this work, we identify ICMEs mainly based on the probable ICME list of \textit{Voyager} 2 given by \citet{Wang and Richardson 2004}. In order to insure the availability of combined magnetic field and plasma data with temporal resolution of 48 s or 96 seconds (which is adequate for analyzing AFs suggested by \citet{Li et al 2016b}) and to avoid the complications caused by the heating of interstellar pickup ions \citep{Richardson and Smith 2003}, we only use \textit{Voyager 2} data from 1977 December 1 to the end of 1979. During this time period, a total of 33 probable ICMEs are identified. Note that, some minor adjustments of ICME boundaries have been done based on the data sets with a higher time resolution compared to hourly data sets used by \citet{Wang and Richardson 2004}.

To reduce the uncertainties of AFs diagnosis introduced in the determinations of the background magnetic field and the deHoffmann-Teller (HT) frame, we apply the approach proposed by \citet{Li et al 2016a} to identify AFs within ICMEs. Instead of the original data sets, the bandpass filtered signals of the plasma velocity and magnetic field observations are used to check the Wal\'en relation as follows:
\begin{linenomath}
\begin{equation}
\label{wr4}
   \delta \mathbf{V}_i = \pm \delta \mathbf{V}_\mathrm{Ai}
\end{equation}
\end{linenomath}
Here, $\delta \mathbf{V}_i$ and $\delta \mathbf{V}_\mathrm{Ai}$ represent the bandpassed $\mathbf{V}$ (solar wind velocity) and $\mathbf{V}_\mathrm{A}$ (local Alfv\'en velocity) with the $i$-$th$ filter, respectively. The $-$/$+$ signs respectively denote the propagation parallell and anti-parallel to the background magnetic field. The parameter proposed by \citet{Li et al 2016a, Li et al 2016b}, $E_{rr}$, is used to assess the goodness of the degree of the Alfv\'enicity. For each filtered data set, we calculate the following eight parameters: 1) $\left|\left|\gamma_c\right|-1\right|$; 2) $\left|\left|\gamma_{cx}\right|-1\right|$; 3) $\left|\left|\gamma_{cy}\right|-1\right|$; 4) $\left|\left|\gamma_{cz}\right|-1\right|$; 5) $\left|\frac{\sigma_{\delta \mathbf{V}}}{\sigma_{\delta\mathbf{V}_\mathrm{A}}}-1\right|$; 6) $\left|\frac{\sigma_{\delta V_x}}{\sigma_{\delta V_{\mathrm{A}x}}}-1\right|$; 7) $\left|\frac{\sigma_{\delta V_y}}{\sigma_{\delta V_{\mathrm{A}y}}}-1\right|$; 8) $\left|\frac{\sigma_{\delta V_z}}{\sigma_{\delta V_{\mathrm{A}z}}}-1\right|$. Here, $\gamma_c$ is the correlation coefficient between all the components of $\delta \mathbf{V}$ and $\delta \mathbf{V}_\mathrm{A}$, $\sigma_{\delta \mathbf{V}}$ represents the standard deviation of all the components of $\delta \mathbf{V}$, and $\sigma_{\delta \mathbf{V}_\mathrm{A}}$ represents the standard deviation of all the components of $\delta \mathbf{V}_\mathrm{A}$. The terms with subscript $x, y,$ and $z$ are for the $x, y$, and $z$ components. The parameter $E_{rr}$ is the average value for these eight parameters. Compared to previous parameters defined to represent the Alfv\'enicity, such as  the Alfv\'en ratio, the Wal\'en slope, the normalized cross helicity, the normalized residual energy, and the velocity-magnetic field correlation coefficient,  $E_{rr}$ is a more comprehensive and reliable quantity \citep{Li et al 2016b}. 

We apply a moving window with a width of 1 hr and a moving step of 5 min to calculate $E_{rr}$ for each filtered data sets. The AFs are defined as the intervals with $E_{rr} < 0.15$ as used by \citet{Li et al 2016a}. For 48 s \textit{Voyager 2} data, the filters are chosen to be 100--135 s, 135--180 s, 180--250 s, 250--330 s, 330--450 s, 450--600 s, 600--810 s, 810--1100 s, 1100--1480 s, and 1480--2000 s. For 96 s \textit{Voyager 2} data, the filters are chosen to be 200--250 s, 250--20 s, 320--400 s, 400--500 s, 500--630 s, 630--800 s, 800--1000 s, 1000--1260 s, 1260--1580 s, and 1580--2000 s.

\begin{figure}[htbp!]
\centering
\noindent\includegraphics[width=30pc]{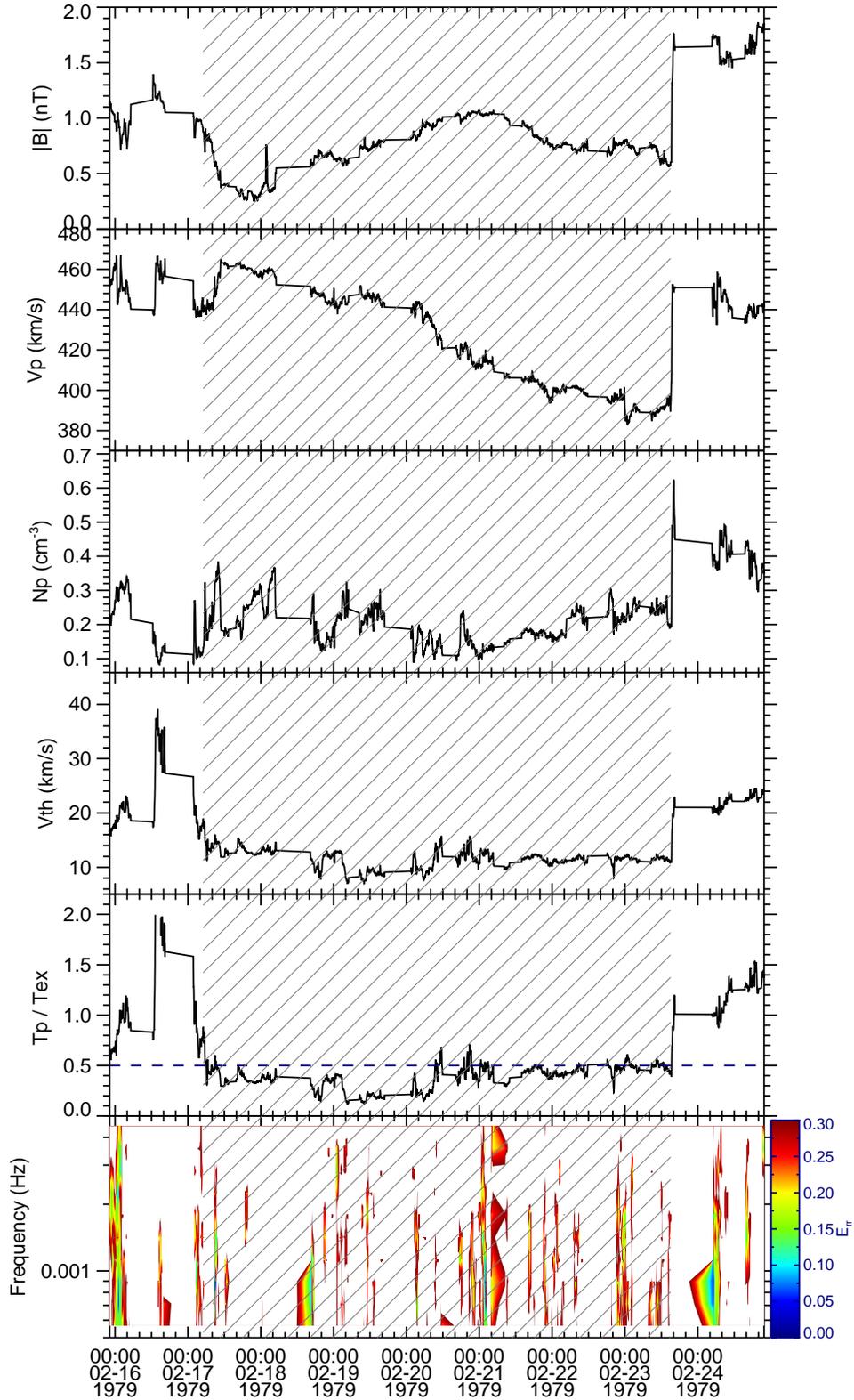}
\caption{Overview of an ICME (hatched area) observed by \textit{Voyager} 2 at $\sim$ 4.73 au. From top to bottom, the panels show the magnetic field strength ($|B|$), the solar wind bulk speed ($V_p$), the proton number density ($N_p$), the proton thermal speed ($V_{th}$), the ratio of the observed to the expected temperature ($T_p/T_{ex}$), and $E_{rr}$, respectively.}
\label{ICME}
\end{figure}

\section{AFs inside an ICME at 4.73 au: a typical case}
\label{event}
Figure \ref{ICME} shows the overview of an ICME observed by \textit{Voyager 2} at $\sim$ 4.73 au during 1979 February 17--23. From top to bottom, the panels show the magnetic field strength ($|B|$), the solar wind bulk speed ($V_p$), the proton number density ($N_p$), the proton thermal speed ($V_{th}$), the ratio of the observed to the expected proton temperature ($T_p/T_{ex}$), and $E_{rr}$, respectively. The threshold value $T_p/T_{ex}$ = 0.5 is plotted as the horizontal dashed line in the fifth panel. As described previously, the primary criteria for identifying possible ICMEs, $T_p/T_{ex}$, is well under 0.5 inside the ICME (hatched area). A monotonic declining of solar wind bulk speed and a cool proton thermal speed ($<$ 20 km/s) are other typical characteristics of a candidate ICME event observed beyond 1 au \citep{Russell and Shinde 2003}. For this event, the solar wind bulk speed decreases nearly monotonically across the ICME and the proton thermal speed within the ICME is less than 15 km/s. The speed of the leading edge is 450 km/s, which is faster than that of the trailing edge of 390 km/s. The speed difference of 60 km/s suggests that the ICME is still expanding as it moves outward. This value is a little larger than the average expansion speed of an ICME at 4.73 au (48$\pm$4 km/s) estimated based on the empirical formula given by \citet{Liu et al 2005}. The density inside ICMEs beyond 1 au is often smaller than in the ambient solar wind. For this event, the density is generally $\leq$ 0.3 cm$^{-3}$, less than the value in the ambient solar wind of 0.4 cm$^{-3}$. In addition, the magnetic field strength has an enhancement during this event. These additional signatures give us more confident that this event is an ICME event. The duration of this ICME is about 154.0 hr with an average solar wind speed of about 420 km/s, which gives a radial width of about 1.55 au, which is a little larger than the average radial width of an ICME at 4.73 au (1.16$\pm$0.04 au) estimated based on the empirical formula given by \citet{Liu et al 2005}. The time-frequency distribution of $E_{rr}$ reveals that there exists many intervals of relatively pure AFs in the center and at both boundaries of the ICME, which are denoted by the green and blue regions.

\begin{figure}
\centering
\noindent\includegraphics[width=20pc]{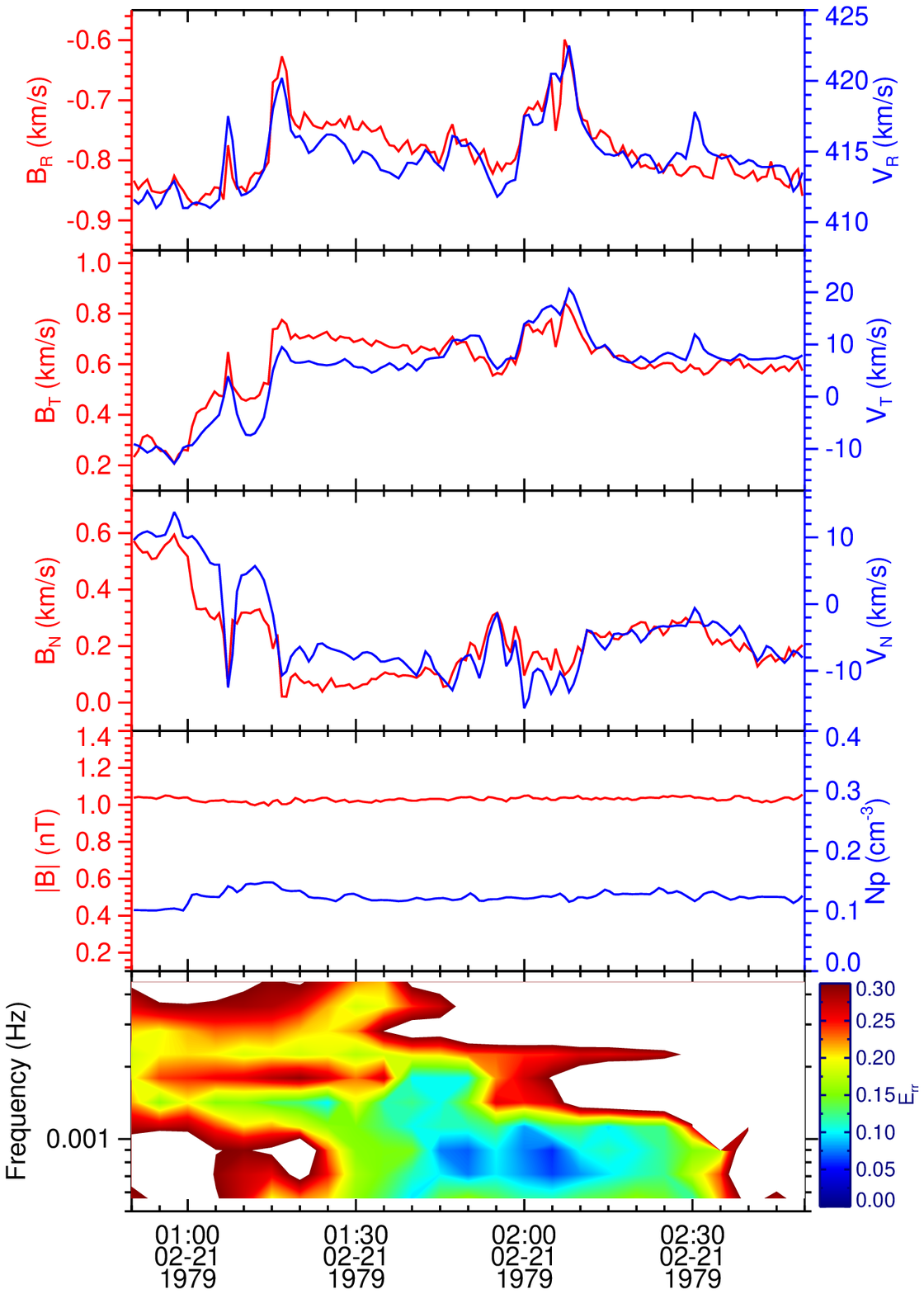}
\noindent\includegraphics[width=20pc]{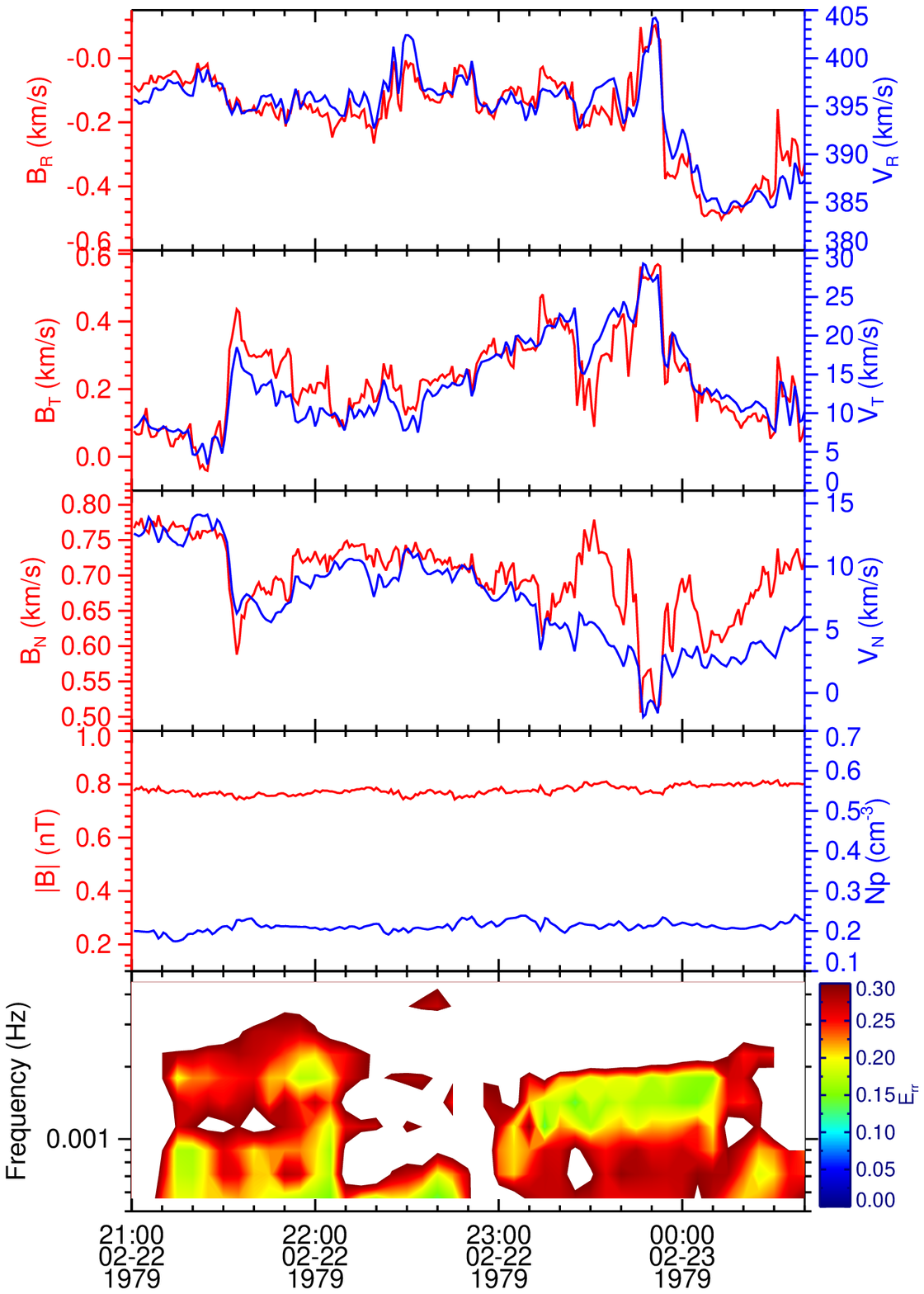}
\caption{Two AFs inside the ICME shown in Figure \ref{ICME}. The left: 0050--0250 UT on 1979 February 21; the right: 2100 UT on 1979 February 22 to 0040 UT on 1979 February 23. The first three panels show the magnetic field ($\mathbf{B}$, in red) and solar wind velocity ($\mathbf{V}$, in blue) in the RTN coordinates, and the fourth panels shows the magnetic field strength ($|B|$, in red) and proton number density ($N_p$, in blue), and the bottom panel shows $E_{rr}$.}
\label{AFEvent}
\end{figure}

Figure \ref{AFEvent} shows two examples of AFs inside the ICME shown in Figure \ref{ICME}. The first three panels show the magnetic field ($\mathbf{B}$, in red) and solar wind velocity ($\mathbf{V}$, in blue) in the RTN coordinates, the fourth panels shows the magnetic field strength ($|B|$, in red) and proton number density ($N_p$, in blue), and the bottom panel shows $E_{rr}$. The left figure shows the AFs during  0050--0250 UT on 1979 February 21. During this time interval, the solar wind is essentially incompressible with relative fluctuations $\delta N_p/N_p$ of 7.7\% and $\delta |B|/|B|$ of 1.1\%. However, the three components of $\mathbf{B}$ and $\mathbf{V}$ have large-amplitude fluctuations which have a strong positive correlation. The correlation coefficients for the R, T, and N components are 0.84, 0.89, and 0.88, respectively. Such a strong correlation and incompressibility indicate the presence of AFs propagating anti-parallel to the background magnetic field, which is assumed to be the mean magnetic field during this time interval (--0.79, 0.60, 0.22) nT. From the time-frequency distribution of $E_{rr}$, it is clear that the AFs during this interval are not periodic like a monochromatic wave, instead, are broadband with different frequency at different time. For example, the wave periods of relatively pure AF during 0050 and 0150 UT are generally 630--800 s, and the wave periods during 0135--0230 UT change to 800--2000 s. The right figure shows the AFs from 2100 UT on 1979 February 22 to 0040 UT on 1979 February 23. The relative fluctuations of $\delta N_p/N_p$ and $\delta |B|/|B|$ are insignificant, 5.6\% and 2.2\%, respectively. This indicates that the solar wind is incompressible. However, the fluctuations in the three components of $\mathbf{B}$ and $\mathbf{V}$ in the RTN coordinates have large amplitudes and strong positive correlation. The correlation coefficients for the R, T, and N components are 0.93, 0.88, and 0.82, respectively. The background magnetic field, which is assumed to be the mean magnetic field during this time interval, is (--0.17, 0.22, 0.70) nT. Obviously, there exists relatively pure AFs during this time interval, and propagate anti-parallel to the N axis in the RTN coordinates. This is confirmed from the time-frequency distribution of $E_{rr}$ as well. The wave periods of relatively pure AF during 2110 and 2210 UT are generally 1260--2000 s, and the wave periods during 2310--0010 UT change to 630--1000 s.

\section{Statistical Characteristics of AF inside ICMEs}
\label{stat}
\begin{figure}[htbp!]
\centering
\noindent\includegraphics[width=28pc]{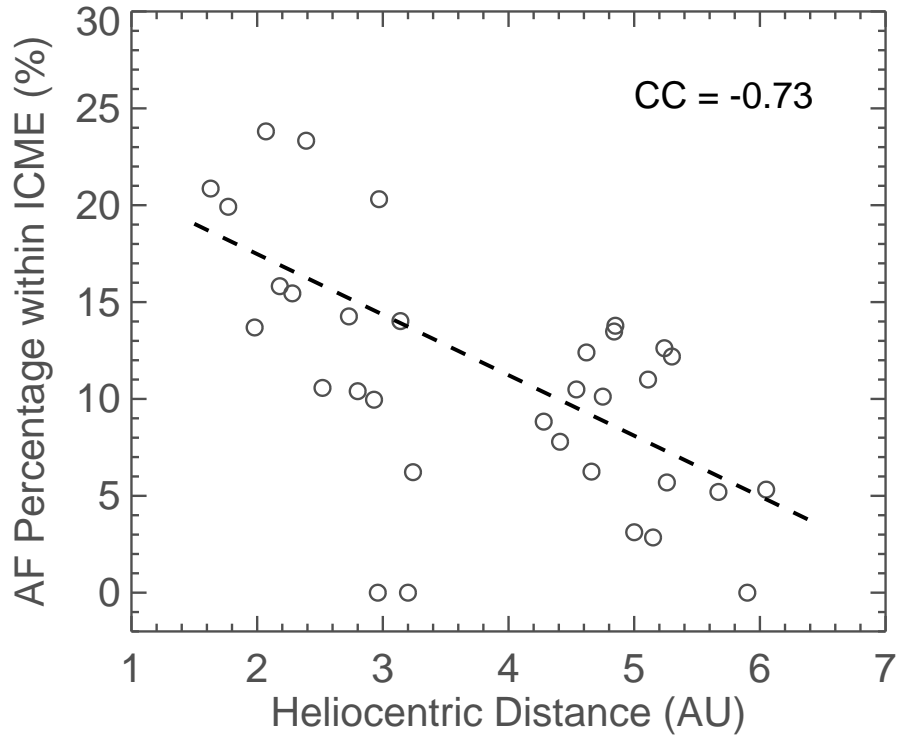}
\caption{Dependence of the percentage of AF duration inside ICME on heliocentric distance. The dashed line denotes the linear fitting result. CC represents the correlation coefficient.}
\label{ratio}
\end{figure}

The results shown in section \ref{event} suggest that relatively pure AFs can and do exist inside ICMEs. Figure \ref{ratio} shows the dependence of the percentage of AF duration inside ICMEs on heliocentric distance. Among 33 probable ICMEs observed by \textit{Voyager} 2 from 1977 December 1 to the end of 1979, AFs could be identified inside 30 ICMEs, 91\% percentage. Besides, it is obvious that the percentage of AF duration reduces generally linearly as ICMEs expand and move outward. The dashed line represents the linear fitting result with the correlation coefficient of --0.73. For ICMEs at $\sim$ 2 au, the percentage of AF is about 20\%, while for ICMEs at $\sim$ 6 au, the percentage of AF decreases significantly to about 5\%.

\floattable
\begin{deluxetable}{ccccc}
\tablecolumns{5}
\tablenum{1}
\tablewidth{0pt}
\tablecaption{AFs inside ICMEs and in ambient solar wind \label{tab1}}
\tablehead{\colhead{} & \colhead{}  & \colhead{1 $\sim$ 4 au} & \colhead{4 $\sim$ 6 au} & \colhead{All events}}
\startdata
{       } & ICME duration\tablenotemark{a} & 488.5 & 1723.2 & 2211.7 \\
ICME & AFs within ICMEs & 72.2 & 180.7 & 252.9 \\
{       } & Percentage & 14.8\% & 10.5\% & 11.4\% \\
\hline
{       } & SW duration\tablenotemark{b} & 3385.2 & 5170.6 & 8555.8 \\
SW & AFs in ambient SW & 762.8 & 608.5 & 1371.3 \\
{       } & Percentage & 22.5\% & 11.8\% & 16.0\%\\
\enddata
{\em      }\tablenotetext{a}{ Data gaps have been removed.}
\tablenotetext{b}{ Data gaps and data during ICMEs have been removed.}
\end{deluxetable}

Table \ref{tab1} gives some statistical characteristics of AF occurrence rate inside ICMEs and in ambient solar wind at different heliocentric distance. For all events, the total duration of AFs with ICMEs is 252.9 hr, about 11.4\% of the total ICME duration (2211.7 hrs) with data gap removed. The occurrence rate of AFs decreases as ICMEs expand and move outward. For AFs within ICMEs at between 1 $\sim$ 4 au, the total duration is 72.2 hr, about 14.8\% of the total ICME duration (488.5 hr). For AFs within ICMEs at between 4 $\sim$ 6 au, the total duration is 180.7 hr, about 10.5\% of the total ICME duration (1723.2 hr). In general, the occurrence rate of AFs inside ICMEs is found to be much less than that in ambient solar wind at 1 au \citep{Liu et al 2006}. For all events, the total duration of AFs in ambient solar wind is 1371.3 hr, about 16.0\% of the total duration of solar wind observation with data gaps and ICMEs removed. For AFs in ambient solar wind at between 1 $\sim$ 4 au, the total duration is 762.8 hr, about 22.5\% of the total solar wind duration (3385.2 hr), which is indeed much larger than that inside ICMEs and in consistent with previous findings. For AFs in ambient solar wind at between 4 $\sim$ 6 au, the total duration is 608.5 hr, about 11.8\% of the total solar wind duration (3385.2 hr), which is similar to that inside ICMEs.

\begin{figure}[htbp!]
\centering
\noindent\includegraphics[width=28pc]{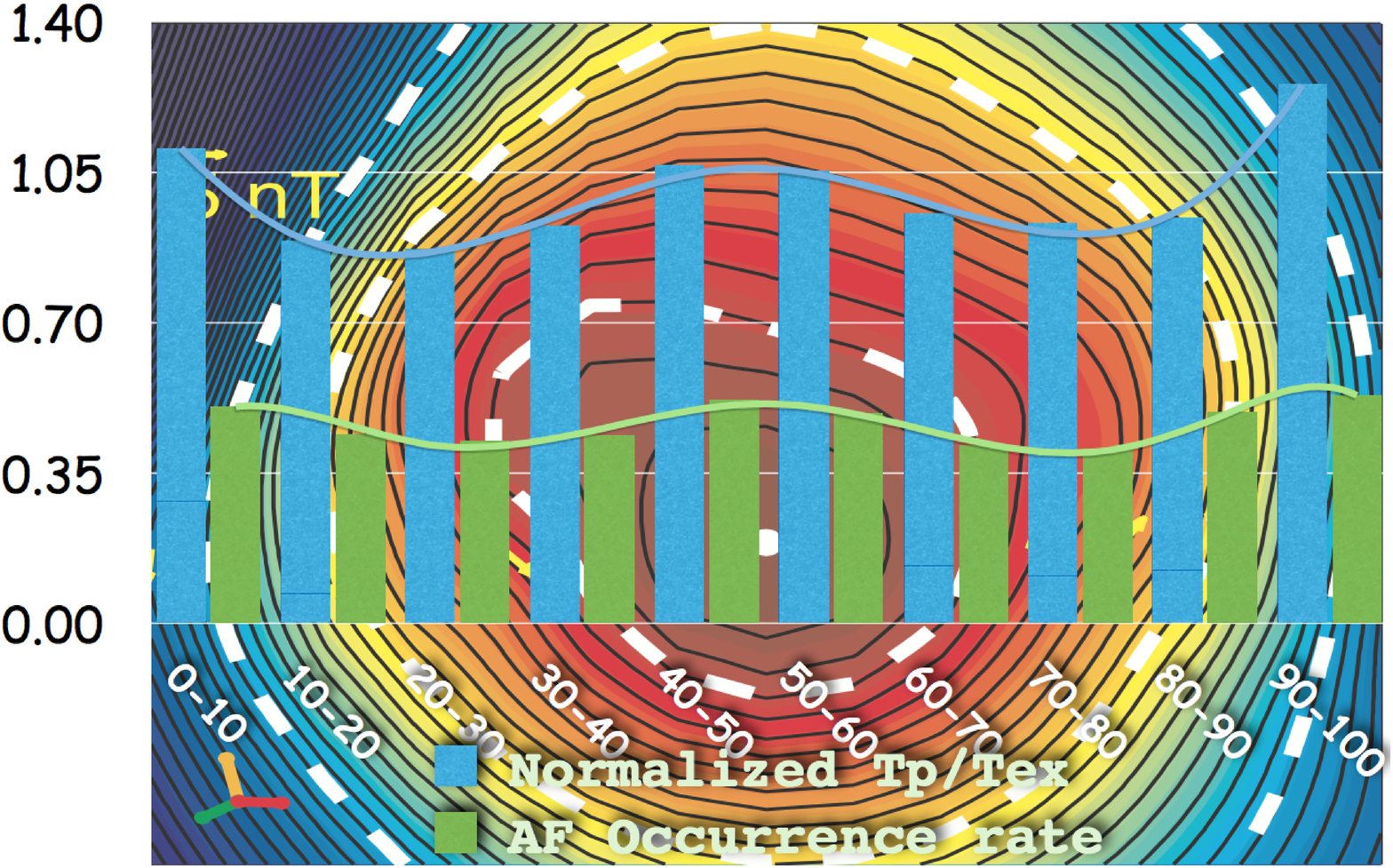}
\noindent\includegraphics[width=28pc]{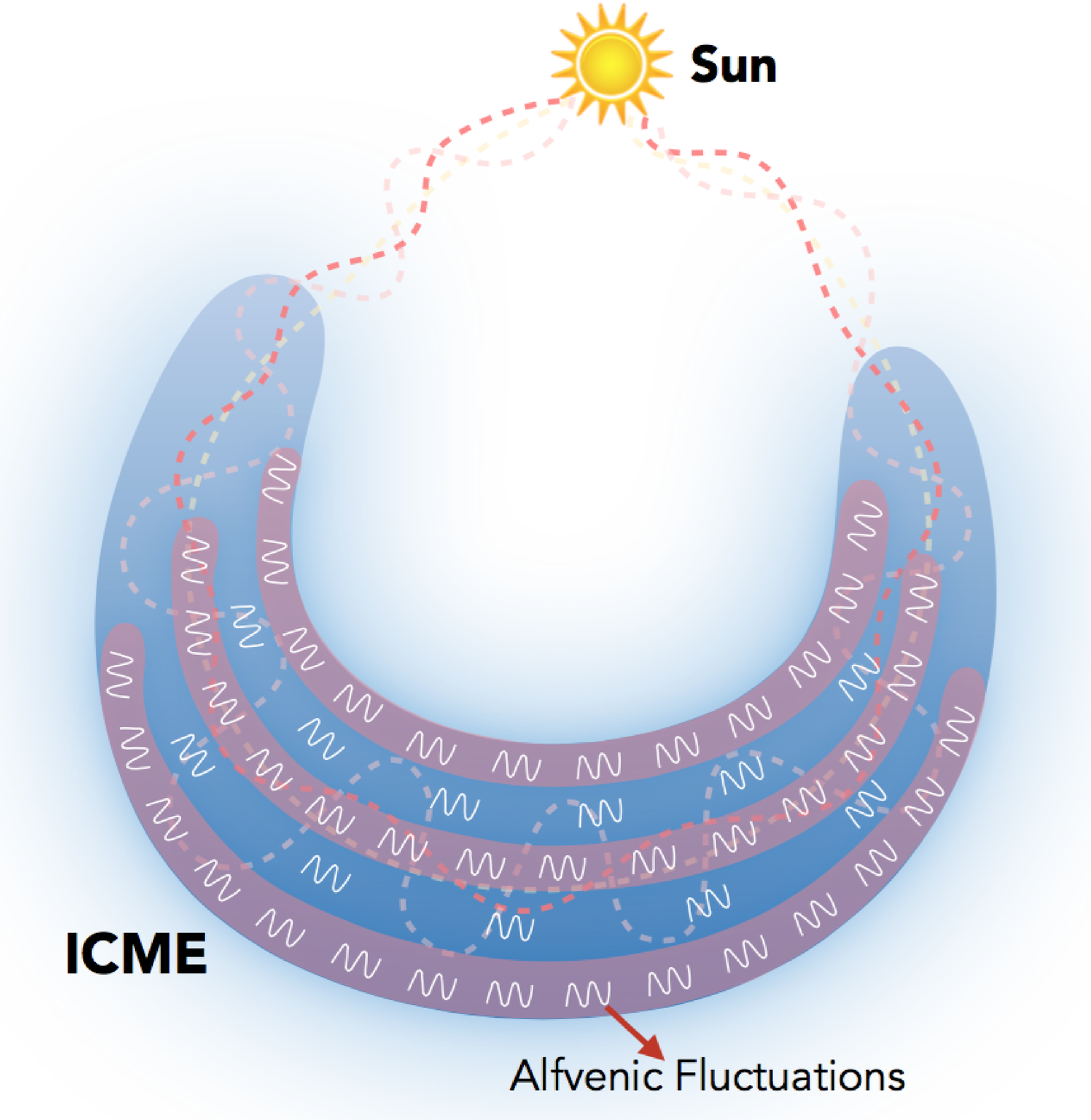}
\caption{Top: Distribution of AF occurrence rate and normalized $T_p/T_{ex}$ across ICMEs based on the superposed epoch analysis. Note that the AF occurrence rate have been multiplied by 1.5 and then plus 0.3  to be shown better together with normalized $T_p/T_{ex}$. The background figure, as adopted from Figure 9 in \citet{Hu and Sonnerup 2002}, is used to illustrate the spacecraft trajectory through the magnetic cloud. Bottom: A sketch of ICME plasma heating due to AF dissipation. The white wave-like curve represents AF. Its spacing indicates the occurrent rate of AF inside ICMEs. The color represent the plasma temperature. The light blue denotes cold and the light red denotes warm.}
\label{dis}
\end{figure}

\section{Indirect evidence of local ICME plasma heating by AF dissipation}
By using the new approach of AF diagnosis, many intervals of AFs are identified inside ICMEs from 1 to 6 au. In addition, the percentage of AF duration inside ICME decreases generally linearly with heliocentric distance. Based on these two findings, we are more confident that AFs dissipation inside ICMEs could contribution to local ICME plasma heating. Here we will show some indirect evidence to link local ICME plasma heating with AFs inside ICMEs. We divide each ICME duration into 10 segments and obtain the distribution of AF occurrence rate and normalized $T_p/T_{ex}$ across ICMEs based on the superposed epoch analysis, shown in the top of Figure \ref{dis}. Note that the AF occurrence rate have been multiplied by 1.5 and then plus 0.3 in order to be better shown together with normalized $T_p/T_{ex}$. The horizontal axis represents the relative location across the ICME cross-section. 0--10 denotes the lading edge of an ICME, 90--100 denotes the trailing edge of an ICME, and 40--60 denotes the center of an ICME. The green bars show the distribution of AF occurrence rate inside ICMEs, and the green line represents the polynomial fitting result. The distribution of AF occurrence rate inside ICMEs represents a clear ``W" shape, indicating that the AFs are more frequently found in the center and at the boundaries of ICMEs. The blue bars give the distribution of normalized $T_p/T_{ex}$, and the blue line represents the polynomial fitting result. Similarly, the ``W" shaped distribution is obviously found in normalized $T_p/T_{ex}$, indicating that the ICME plasma seems to be more heated in the center and at the boundaries of ICMEs.

To interpret these phenomena physically, some assumptions meed to be made in advance: 1) The temperature inside ICMEs is nearly uniform at the beginning; 2) The whole ICME structure experiences the same expansion; 3) AFs inside ICMEs originate from the Sun's surface when the CME occurs; 4) AF distribution inside ICMEs is nonuniform; 5) the dissipation rates of AF inside ICMEs are identical. The bottom of Figure \ref{dis} gives a sketch of ICME plasma heating due to AF dissipation. AFs are more frequently found in the center and at the boundaries of ICMEs. Considering a nearly identical dissipation rate, more AFs would be dissipated in the center and at the boundaries of ICMEs, whereafter more energy would contribute to ICME plasma heating in the center and at the boundaries. As ICMEs expand and propagate outward, the percentage of AF duration inside ICMEs keeps on decreasing with heliocentric distance, which has been confirmed in section \ref{stat} .

\section{Summary}
Nonlinear cascade of low-frequency Alfv\'enic fluctuations (AFs) is regarded as one of the major candidate mechanisms of local ICME plasma heating during its expansion and transportation. However, AFs  inside ICMEs have been rarely reported in the literature. In this study, we identify 33 probable ICMEs observed by \textit{Voyager} 2 between 1 and 6 au, finding that relatively pure AFs could be frequently seen inside 30 ICMEs with an average occurrence rate of 12.6\%. Statistically, the percentage of AF duration inside ICMEs decays generally linearly as ICMEs expand and move outward. Compared to in ambient solar wind, the occurrence rate of AFs inside ICMEs is much less, especially within 4 au. Furthermore, the occurrence rate of AFs and the proton temperature inside ICMEs have similar ``W"-shaped distributions, large in the center and at the boundaries of ICMEs. By assuming an uniform dissipation rate of AFs inside ICMEs, our findings provide an indirect evidence of local ICME plasma heating due to AF dissipation.

\acknowledgments

We thank the \emph{Voyager} Mission for the use of their data. The high-resolution \emph{Voyager 2} plasma data are publicly available at the \emph{Voyager} Data Page of the Massachusetts Institute of Technology Space Plasma Group (\url{ftp://space.mit.edu/pub/plasma/vgr/v2}). The 48 s resolution magnetic field data are accessible at the NSSDC (\url{ftp://nssdcftp.gsfc.nasa.gov}). The authors would like to thank Dr. Shuo Yao for helpful discussions. This work was supported by 973 program 2012CB825602, NNSFC grants 41574169, 41574168. H. Li was also supported by Youth Innovation Promotion Association of the Chinese Academy of Sciences and in part by the Specialized Research Fund for State Key Laboratories of China.


\begin{thebibliography}{}

\bibitem[Bemporad et al.(2007)]{Bemporad et al 2007}
Bemporad, A., Raymond, J., Poletto, G., \& Romoli, M. 2007, \apj, 655, 576

\bibitem[Burlaga(1995)]{Burlaga 1995}
Burlaga, L. F.\ 1995, Interplanetary Magnetohydrodynamics (New York, Oxford University Press)

\bibitem[Filippov \& Koutchmy(2002)]{Filippov and Koutchmy 2002}
Filippov, B., \& Koutchmy, S.\ 2002, \solphys, 208, 283

\bibitem[Furth et al.(1963)]{Furth et al 1963}
Furth, H. P., Killeen, J., \& Rosenbluth, M. N.\ 1963, Phys. Fluids, 6, 459

\bibitem[Galinsky \& Shevchenko(2012)]{Galinsky and Shevchenko 2012}
Galinsky, V. L., \& Shevchenko, V. I.\ 2012, \apj, 751, 146

\bibitem[Gosling(1990)]{Gosling 1990}
Gosling, J. T.\ 1990, in Physics of Magnetic Flux Ropes, ed. C. T. Russell, E. R. Priest, \& L. C. Lee (Geophys. Monogr. Ser. 58; Washington, DC: AGU), 343

\bibitem[Kasper et al.(2008)]{Kasper et al 2008}
Kasper, J. C., Lazarus, A. J., \& Gary, S. P.\ 2008, \prl, 101, 261103


\bibitem[Hu \& Sonnerup(2002)]{Hu and Sonnerup 2002}
Hu, Q., \& Sonnerup, B. U. \"{O},\ 2002, \jgr, 107, 1142

\bibitem[Landi et al.(2010)]{Landi et al 2010}
Landi, E., Raymond, J. C., Miralles, M. P., \& Hara, H.\ 2010, \apj, 711, 75

\bibitem[Li et al.(2016a)]{Li et al 2016a}
Li, H., Wang, C., Chao, J. K., \& Hsieh, W. C. \ 2016a, \jgr, 121, 42

\bibitem[Li et al.(2016b)]{Li et al 2016b}
Li, H., Wang, C., Belcher, J. W., He, J. S., \& Richardson, J. D.\ 2016b, \apjl, 824, L2

\bibitem[Liu et al.(2005)]{Liu et al 2005}
Liu, Y.,  Richardson J. D. , \& Belcher J. W.\ 2005, \planss, 53, 3

\bibitem[Liu et al.(2006)]{Liu et al 2006}
Liu, Y.,  Richardson J. D. , Belcher J. W., \& Kasper, J. C.\ 2006, \jgr, 111, A01102

\bibitem[Lopez(1987)]{Lopez 1987}
Lopez, R. E.\ 1987, \jgr, 92, 11189

\bibitem[Marsch et al.(2009)]{Marsch et al 2009}
Marsch, E., Yao, S., \& Tu, C. -Y.\ 2009, Ann. Geophys., 27, 869

\bibitem[Murphy et al.(2011)]{Murphy et al 2011}
Murphy, N. A., Raymond, J. C., \& Korreck K. E.\ 2011, \apj, 735, 17

\bibitem[Neugebauer \& Goldstein(1997)]{Neugebauer and Goldstein 1997}
Neugebauer, M., \& Goldstein, R.\ 1997, in Coronal Mass Ejections, ed. N. Crooker, J. A. Joselyn, \& J. Feynman (Geophys. Monogr. Ser. 99; Washington, DC: AGU), 245

\bibitem[Richardson \& Cane(1995)]{Richardson and Cane 1995}
Richardson, I. G., \& Cane, H. V.\ 1995, \jgr, 100, 23397

\bibitem[Richardson \& Smith(2003)]{Richardson and Smith 2003}
Richardson, J. D., \& Smith, C. W.\ 2003, \grl, 30, 1206

\bibitem[Richardson et al.(2006)]{Richardson et al 2006}
Richardson, J. D., Liu, Y., Wang, C., \& Burlaga, L. F.\ 2006, Adv. in Space Res., 38, 528

\bibitem[Russell \& Shinde(2003)]{Russell and Shinde 2003}
Russell, C. T., \& Shinde, A. A.\ 2003, \ssr, 216, 285

\bibitem[Rust \& LaBonte(2005)]{Rust and LaBonte 2005}
Rust, D. M., \& LaBonte, B. J.\ 2005, \apj, 622, L69

\bibitem[Tam \& Chang(1999)]{Tam and Chang 1999}
Tam, S. W. Y., \& Chang, T. 1999, \grl, 26, 3189

\bibitem[Tu \& Marsch(1995)]{Tu and Marsch 1995}
Tu, C. -Y., \& Marsch, E.\ 1995, \ssr, 73, 1

\bibitem[Wang \& Richardson(2004)]{Wang and Richardson 2004}
Wang, C., \& Richardson, J. D.\ 2004, \jgr, 109, A06104

\bibitem[Wang et al.(2005)]{Wang et al 2005}
Wang, C., Du D., \& Richardson, J. D.\ 2005, \jgr, 110, A10107

\bibitem[Wang et al.(2014)]{Wang et al 2014}
Wang, C. B., Wang, B., \& Lee, L. C.\ 2014, \solphys, 289, 3895

\bibitem[Yao et al.(2010)]{Yao et al 2010}
Yao, S., Marsch, E., Tu, C. -Y., \& Schwenn, R.\ 2010, \jgr, 115, A05103

\bibitem[Zurbuchen \& Richardson(2006)]{Zurbuchen and Richardson 2006}
Zurbuchen, T. H., \& Richardson, I. G.\ 2006, \ssr, 123, 31


\end{thebibliography}
\end{document}